\title{Quantum Divide-and-Conquer Anchoring for Separable Non-negative Matrix Factorization}
\author{%
	Yuxuan Du\footremember{alley}{UBTECH Sydney AI Centre and the School of Information Technologies in the Faculty Engineering and
		Information Technologies at The University of Sydney, NSW, 2006, Australia, yudu5543@uni.sydney.edu.au,
		tongliang.liu@sydney.edu.au, dacheng.tao@sydney.edu.au }%
	\and Tongliang Liu \footrecall{alley}%
	\and Yinan Li \footremember{trailer}{Centre for Quantum Software and Information,
		Faculty of Engineering and Information Technology,
		University of Technology Sydney, NSW 2007, Australia, yinan.li@student.uts.edu.au, runyao.duan@uts.edu.au} 
	\and Runyao Duan\footrecall{trailer} \footnote{UTS-AMSS Joint Research Laboratory for Quantum Computation and Quantum Information Processing, Academy of Mathematics and Systems Science, Chinese Academy of Sciences, Beijing 100190, China}
	\and Dacheng Tao\footrecall{alley}  
}
\date{}

\documentclass[12pt]{article}
\usepackage{amsmath}
\usepackage{times}
\usepackage{xcolor}
\usepackage{soul}
\usepackage[utf8]{inputenc}
\usepackage[small]{caption}
\usepackage{amsfonts}
\usepackage{amsthm}
\usepackage{amsfonts,amssymb,amsthm}
\usepackage[colorlinks]{hyperref}
\usepackage{braket}

\usepackage{algorithmic}
\usepackage[linesnumbered,ruled]{algorithm2e}
\usepackage{graphicx}
\usepackage[toc,page]{appendix}
\usepackage[caption = false]{subfig}
\usepackage{natbib}
\newcommand{\C}{{\mathbb{C}}}

\newcommand{\R}{{\mathbb{R}}}

\newtheorem{thm}{Theorem}

\newcommand{\proj}[1]{\ket{#1}\!\bra{#1}}

\usepackage{hyperref}
\newcommand{\footremember}[2]{%
	\footnote{#2}
	\newcounter{#1}
	\setcounter{#1}{\value{footnote}}%
}
\newcommand{\footrecall}[1]{%
	\footnotemark[\value{#1}]%
} 
\begin{document}
	\maketitle
	
	\begin{abstract}
		It is NP-complete to find non-negative factors $W$ and $H$ with fixed rank $r$ from a non-negative matrix $X$ by minimizing $\|X-WH^\top\|_F^2$. Although the separability assumption (all data points are in the conical hull of the extreme rows) enables polynomial-time algorithms, the computational cost is not affordable for big data. This paper investigates how the power of quantum computation can be capitalized to solve the non-negative matrix factorization with the separability assumption (SNMF) by devising a quantum algorithm based on the divide-and-conquer anchoring (DCA) scheme \cite{zhou2013divide}. The design of quantum DCA (QDCA) is challenging. In the divide step,  the random projections in  DCA is completed by a quantum algorithm for linear operations, which achieves the exponential speedup. We then  devise a \textit{heuristic post-selection} procedure which extracts the information of anchors stored in the quantum states efficiently. Under a plausible assumption, QDCA performs efficiently, achieves the quantum speedup, and is beneficial for high dimensional problems.
	\end{abstract}
	
	\section{Introduction} \label{sec:Intro}
Non-negative matrix factorization (NMF) \cite{lee1999learning,pauca2004text} is popular in computer vision and machine learning, because the underlying non-negativity constraints on the two low-rank factors usually yield sparse representations of the given non-negative matrix. It has proven that NMF is NP-complete \cite{vavasis2009complexity}. Thus, the separability assumption has been introduced \cite{donoho2004does} to NMF and induces SNMF. This assumption enables not only polynomial-time algorithms \cite{zhou2013divide,recht2012factoring,NIPS2017_6847}, but also a geometric interpretation \cite{donoho2004does}.	

However, the rapid progress of the Internet technology, and the computational power and storage, as well as the wide distribute of sensors, grows data exponentially, which challenges the polynomial algorithms. Thanks to quantum physics, quantum computing machinery and quantum machine learning are arising \cite{biamonte2017quantum}. Many encouraging results have been reported recently, such as quantum support vector machine \cite{rebentrost2014quantum} and quantum perceptron \cite{kapoor2016quantum},
which dramatically reduce the runtime complexity and achieve a more efficient learning ability.

Through exploiting quantum advantages, a logarithmic runtime complexity of SNMF is desired and then many emergent applications can be beneficial from this acceleration. Thus, we need to consider restrictions in quantum computing and answer the following questions: (1) how to convert the classical SNMF problem to accord with a quantum framework; (2) how to exploit quantum advantages, preferring to achieve the exponential speedup; and (3) how to circumvent reading out bottleneck in measurements \cite{aaronson2015read}.

We select the divide-and-conquer anchoring (DCA) \cite{zhou2013divide} scheme and devise quantum DCA (QDCA) for SNMF, because DCA only contains linear operations in the time-consuming divide step and this characteristic echoes with the nature of quantum computing. This answers the first question and confirms the second question by guaranteeing the exponential speedup for operations in a quantum machine. 

Thanks to that indexes of anchors can be sampled with a high probability from a probability distribution in the resulting quantum states, we propose an efficient \textit{heuristic post-selection} method instead of reading out all the quantum data directly (reading is an expensive operation, especially for high-dimensional vectors). This \textit{heuristic post-selection} method answers the third question and ensures us to achieve the quantum speedup after measurements. 

The exponential speedup for computations in quantum machine and that for transmitting indexes of anchors from quantum machine to classical computer after measurements together guarantee that the runtime complexity of QDCA achieves $O(poly\log(n + m))$, where $n\times m$ is the size of the input non-negative matrix.

In addition, we deliver an important message through this paper. QDCA is the first algorithm that seamlessly integrates quantum and classical computations. Such kind of integration forms a general strategy to develop algorithms with quantum advantages.

The rest of this paper is organized as follows: Section \ref{sec:preliminary} reviews SNMF and DCA; Section \ref{sec:QuantumNMF} elaborates QDCA and analyzes the runtime complexity; and Section \ref{sec:conclusion} concludes the paper and discusses the future works.

\section{Background}\label{sec:preliminary}
NMF aims to approximate a non-negative matrix $X\in \mathbb{R}_+^{n\times m}$ by the product of two nonnegative low rank factors (a basis matrix $W\in \mathbb{R}_+^{n\times r}$ and an encoding matrix $H\in \mathbb{R}_+^{m\times r}$), i.e., $X\approx {W}{H}^\top$, where   $r=O(\log{(n+m)})\ll \min\{n,m\}$, via solving the following optimization problem
\begin{equation}\label{eqn:NMF}
\begin{aligned}
&\min_{W\in \mathbb{R}_+^{n\times r},H\in \mathbb{R}_+^{m\times r}} \frac{1}{2}\|X-WH^\top\|_F^2\\=&\min_{W\in \mathbb{R}_+^{n\times r},H\in \mathbb{R}_+^{m\times r}} \frac{1}{2}\sum_{i=1}^n\sum_{j=1}^{m}(X_{ij}-(WH^\top)_{ij})^2~.
\end{aligned}
\end{equation}

\subsection{Separable Non-negative Matrix Factorization}\label{def:SNMF}
Solving the original NMF problem (equation (\ref{eqn:NMF})) is in general NP-complete~\cite{vavasis2009complexity}. Thus it is computationally intractable to obtain the global optimum in polynomial time with respect to the input size.  To circumvent this difficulty, \cite{donoho2004does} introduced the \emph{separability assumption} to the non-negative matrix $X$, i.e., $X$ can be decomposed into $X = FX(R,:)$, where the basis matrix $X(R,:)$ is composited by $r$ rows from $X$ and $F$ is the non-negative encoding matrix. 

We denote  $R=\{k_1,k_2,...,k_r\}$, where $k_i\in\{1,2,...,n\}$ and $|R|=r$. If a cone  can be defined as $cone(X(R,:)) = \sum_{i=1}^r\alpha_iX({k_i},:),~\alpha_i\in\mathbb{R}_+$, the $cone(X(R,:))$ is the conical hull of $X(R,:)$.
We say $X$ is separable if $\forall k_i\in \{1,...,n\}$, we have 
\begin{equation}
X(R,:)=\{X(k_i,:)\}_{k_i\in R}~,
\end{equation}
where $X(k_i,:)\in cone(X(R,:))$.

\begin{figure}[t!]
	\centering
	\includegraphics[bb=5bp 40bp 1000bp 340bp,scale=0.50]{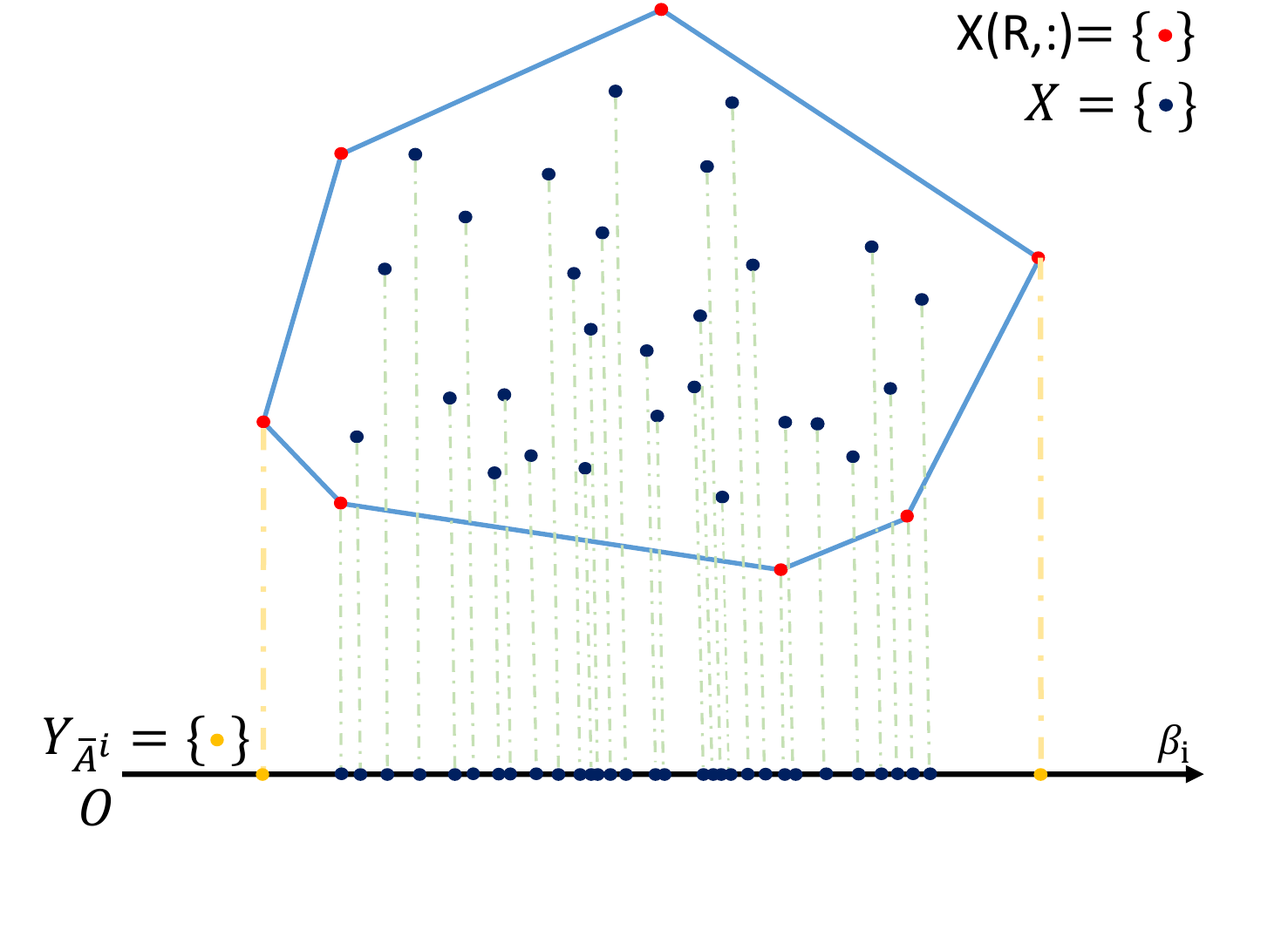}
	
	\caption{\small{ \textit{An illustration for SNMF. The red points stand for the anchors. All data of $X$, except for anchors, are denoted as blue points and are contained in the convex hull, generated by the anchors. After any random projection into $1$-dimensional space, the geometric information is still partially preserved, where the anchors in the projected space are denoted as yellow points.}}}
	\label{fig:cone}
\end{figure}

By adding an extra constraint $\sum_{i=1}^r\alpha_i=1$, we say the simplex $\Delta(X(R,:))$ is the convex hull of $X(R,:)$. This is valuable in practice. 
The selected rows in $X(R,:)$ are called \emph{anchors} (or extreme rows) of  $X$.  The non-anchor vectors in $X$, which are the rest $n-r$ points in $\R^m_+$, lie in the convex hull or conical hull,  generated by the anchors and thus can be non-negatively and linearly expressed by the anchors. Figure \ref{fig:cone} shows the geometrical interpretation of convex hull.  
Likewise, for the near-separable case, all data points are in or around the conical hull of the extreme rows. The concept of near-separable NMF can be straightforwardly  defined by $X=FX(R,:)+N$, where $N$ is a noise matrix for convenient reconstruction.

\subsection{Divide-and-Conquer Anchoring}\label{sec:DCA}
Based on a divide-and-conquer scheme that exploits the geometric information of convex hull or conical hull partially preserved in their projections, divide-and-conquer anchoring (DCA)~\cite{zhou2013divide}  {selects the indexes of anchors from a number of operations in a low-dimensional space, for example the $1$-dimensional space} used in the rest of the paper. DCA is comprised of two parts: (1) the divide step, targeting the collection of indexes of anchors in the $1$-dimensional space, and (2) the conquer step, aiming to determine $R$ via collecting all (for separable case) or high frequency (for near-separable case) indexes of anchors in the $1$-dimensional space. To better understand the proposed QDCA, we detail the procedure of DCA. 

\underline{Divide step.} Given a set of unit vectors $B=\{\beta_i\}_{i=1}^s\in \mathbb{R}^{m\times s}$ randomly sampled from the unit hypersphere $\mathbb{S}^{m-1}$,
where $s=O(r\log r)$,  we project $X$   onto $\beta_i$  and obtain 	 
\begin{equation}\label{def:Y}
Y_i=X\beta_i,~Y_i\in\R^{n}~.
\end{equation}
Denote the indexes of the smallest and the largest entries of $Y_i$ in the $1$-dimensional space as $\bar{A}^i$, i.e.,  
\begin{equation}\label{eqn:1D}
\bar{A}^i:=\left\{\arg\max_{k_j}X(k_j,:)\beta_i,~ \arg\min_{k_j}X(k_j,:)\beta_i\right\}~,
\end{equation}
where $i=\{1,\dots,s\}$. As illustrated in Figure \ref{fig:cone}, each $\bar{A}^i$
corresponds to a particular pair of anchors in the original high dimensional space.

\underline{Conquer step.} 
Under the separability assumption, we find $r$ distinct indexes to identify all the anchors for $X$. Under the near-separability assumption, the indexes of anchors in randomly projected spaces are collected by selecting the most $r$ frequently appeared indexes from $\{\bar{A}^i\}_{i=1}^s$. The selection rule is defined as
\begin{equation}\label{def:conquer}
R:= \arg \max_{H\subseteq[n],|H|=r}\sum_{i\in H}\sum_{j=1}^sI(i\in \bar{A}^j)~,
\end{equation}
where $|H|$ is the size of the set $H$, and $I(i\in \bar{A}^j):~i\rightarrow\{0,1\}$ is the indicator function for the event that an index $i$ is within $\bar{A}^j$ of the $j$-th random projection operation.

\section{Quantum Divide-and-Conquer Anchoring}\label{sec:QuantumNMF}

Devising QDCA is challenging and so non-trivial. To realize quantum advantages for solving SNMF, we shall transform  the DCA scheme. For simplicity, we decompose DCA into $s$ sub-problems corresponding to $s$ random projections, in which the $i$-th sub-problem is comprised of the $i$-th random projection and the subsequent procedure for determining $\bar{A}^i$. It is worth noting that, in QDCA, for preserving the exponential speedup, only the index of an anchor with the maximum absolute value is collected in $\bar{A}^i$, i.e., $\bar{A}^i=\arg \max_{k_j}\{|X(k_j,:)\beta_i|\}$. {The random projection can be completed by an efficient quantum algorithm for linear operations \cite{lloyd2014quantum}, achieving an exponential speed-up with respect to the classical counterpart.} After the random projections, the resulting vectors  will be presented as quantum states  (proportional to $X\beta_i$), which are infeasible to readout all its probability amplitudes in  a logarithmic runtime. To overcome this barrier and target exponential speedup, we devise  a \emph{heuristic post-selection} method to obtain $\{\bar{A}^i\}_{i=1}^s$. In the conquer step, we employ equation (\ref{def:conquer}) in  classical computing. Figure \ref{fig:circuit} shows the diagram of QDCA.

Prior to detail the proposed QDCA, we introduce the notations,  which are necessary to explain our results. The \emph{Dirac notations} are used to follow the convention. Basically, we use $\ket{\psi}\in\mathcal{H}_d$, a  normalized $d$-dimensional vector, to denote a $d$-dimensional pure quantum state in the underlying state (Hilbert) space $\mathcal{H}_d$. We use $\bra{\psi}$ to denote the conjugate transpose of $\ket{\psi}$, i.e., $\ket{\psi}=\bra{\psi^{\dagger}}$. The standard inner product of $\ket{\psi}$ and $\ket{\phi}$ is denoted as $\braket{\psi|\phi}$. For a quantum system with Hamiltonian $H$ (a hermitian matrix, satisfying $H^\dagger=H$), the unitary time evolution is characterized by the \emph{matrix exponential}, $e^{-iHt}:=\sum_{k=0}^{\infty} \frac{1}{k!}(-iH)^kt^k~$,
where $t$ is the simulation time. To simulate a quantum system, we are required to implement a \emph{quantum circuit} which mimics the time evolution $e^{-iHt}$ at any time $t$.
The spectral decomposition of $H$ is denoted by $\sum_{j}\lambda_j\proj{u_j}$, where $\lambda_j$ and $u_j$ are the eigenvalue and the corresponding eigenvector of $H$, respectively. Specifically, the matrix exponential $e^{H}$ admits the form $\sum_{j} e^{\lambda_j}\proj{u_j}$.  By default, we have $\ket{0}=[1~0]^{\top}$ and $\ket{1}=[0~1]^{\top}$. For $n$-qubits, let $\ket{i}$ be the computational basis, where $\ket{i}\in\{\ket{0}, \ket{1}\}^{\otimes n}$ and $\otimes$ stands for the operation of tensor product.  Detailed basic notations and preliminaries for quantum computation are referring to~\cite{nielsen2002quantum}.

\begin{algorithm}[t]
	%
	
	\SetKwInOut{Input}{Input}
	\SetKwInOut{Output}{Output}
	
	\Input{ $X\in\mathbb{R}_+^{n\times m}$ 
		via oracle access (see Subsection \ref{sec:read_X});
		\\  $s=O(\log(m+n))\in\mathbb{R}_+$\\  $k=O(poly\log(n+m))\in\mathbb{R}_+$.}
	\Output{ The indexes of anchors $R$.}
	\If{$m\neq n$ or $X$ is not hermitian}{$X\leftarrow\begin{pmatrix}0&{X}\\{X}^\dagger&0\end{pmatrix}$;}
	Generating random vectors $\{\beta_i\}_{i=1}^s$ and preparing  corresponding quantum states $\{\ket{\beta_i}\}_{i=1}^s$ (see Subsection \ref{sec:read_X})\;
	Preparing a quantum circuit to simulate the unitary time evolution of $e^{iXt}$ (see Subsection \ref{sec:read_X})\;
	\For{$i<s$}{
		Applying the quantum algorithm for linear operations  to produce $k$ copies of $\ket{\psi_i}$:\\
		~~~~~~~~~~~~~~~$\ket{\psi_i}\propto X\beta_i$ (see Subsection \ref{sec:QNMF})\;
		Measuring $k$ copies of $\ket{\psi_i}$ using the computational basis (see Subsection \ref{sec:measurement})\;
		Setting $\bar{A}^i=\{l\}$, where $l$ is the outcome that appears most frequently in quantum measurements\\
		(see Subsection \ref{sec:measurement})\;
	}
	Constructing $R$ by $\{\bar{A}^i\}_{i=1}^s$,
	$R\leftarrow \arg \max_{H\subseteq[n+m],|H|=r}\sum_{i\in H}\sum_{j=1}^sI(i\in \bar{A}^j)$ (see Subsection \ref{sec:class});
	\caption{QDCA}
	\label{alg:1}
\end{algorithm}

Algorithm \ref{alg:1} summarizes the proposed QDCA algorithm. Here, we discuss its runtime complexity. In this paper, we focus on the dependence of runtime complexity on the parameters $m$ and $n$, which are dimensions of the input matrix. {In total $s$ sub-problems, for the random projection operations, there exists a quantum algorithm which runs in time $O(poly~\log(m+n))$, achieving the exponential speedup with respect to classical counterparts (see Subsections~\ref{sec:read_X} and \ref{sec:QNMF}). Indexes of anchors in projected spaces $\bar{A}^i$ for $i=\{1,...,s\}$ will be obtained by a  newly designed \emph{heuristic post-selection} method. 
	
	Under a plausible assumption, we prove that $O(poly~\log(m+n))$ runtime is sufficient to locate the $s$ indexes corresponding to the largest absolute amplitude of resulting quantum states with a high probability, which constructs $\{\bar{A}^i\}_{i=1}^s$ and maintains the exponential speedup achieved in the previous steps. However, we have not provided a rigorous runtime analysis for the worst case without any assumption, which is an interesting open problem. In the conquer step (Subsection \ref{sec:class}),  a classical sorting algorithm is  employed, which runs in time $O(s\log{s})$, where $s=O(r\log{r})$ and $r=O(\log{(n+m)})$. This implies that the conquer step can be completed with a runtime far less than $O(poly~\log(m+n))$. Under the plausible assumption,  the overall runtime of QDCA is $O(ploy~\log{(m+n)})$.

	In the following subsections, we detail the QDCA.
	Subsection \ref{sec:read_X} introduces how to read classical data into quantum computer. To complete the divide step under a logarithmic runtime, Subsections \ref{sec:QNMF} and \ref{sec:measurement} sequentially demonstrate how to employ the quantum algorithm for linear operations to realize random projections and how to devise \emph{heuristic post-selection} for transmitting. 
	Finally,  Subsection \ref{sec:class} shows that applying a classical sorting algorithm in the conquer step  preserves the logarithmic runtime in QDCA.

	\begin{figure*}
		{\includegraphics[width=6in]{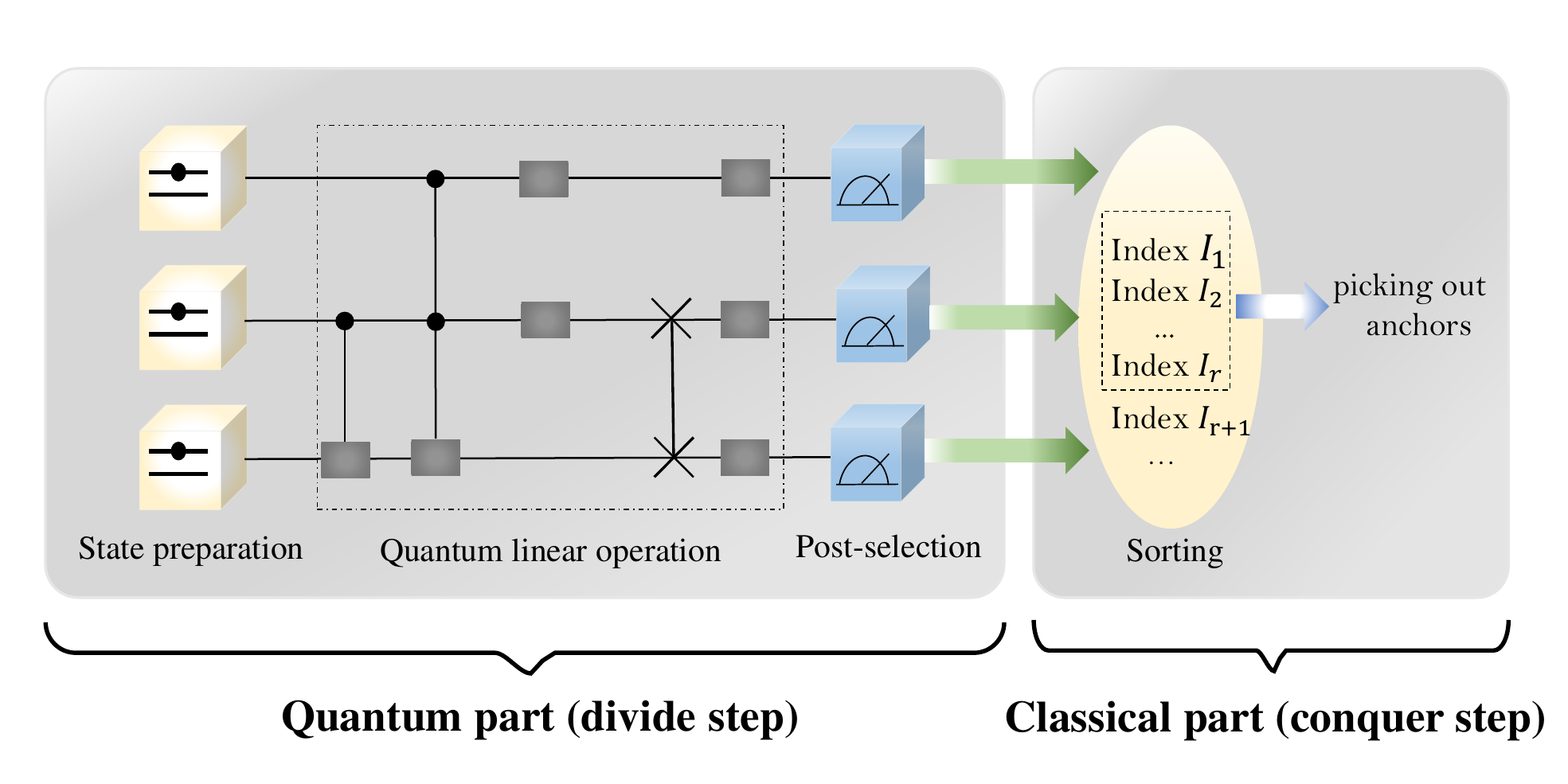}}
		\centering
		\caption{\small{\textit{The circuit of QDCA. The circuit is composed of the quantum part and the classical part. In the quantum part, the classical data are encoded into the quantum form and the random projections are performed through the quantum circuit. Then, the \textit{heuristic post-selection} collects  indexes set $\{\bar{A}^i\}_{i=1}^s$ with the classical form. Finally, the classical part sorting the top $r$ most frequently appearing indexes from $\{\bar{A}^i\}_{i=1}^s$ as the anchor indexes $R$.}}}
		\label{fig:circuit}
	\end{figure*}

	\subsection{Quantum Simulation}\label{sec:read_X}
	For the sake of exploiting quantum advantages, we first describe how to encode the given classical data matrix and random vectors into quantum states.

	Without loss of generality, the input of the SNMF problem is a data matrix $\tilde{X}\in\mathbb{R}^{n\times m}$ with  $rank(\tilde{X})\ll \min\{n,m\}$. In the rest of the paper, we focus on the general case, where $m\neq n$. Referring to \cite{harrow2009quantum},  for satisfying the quantum computing requirement, we should embed $\tilde{X}$ into a high-dimensional matrix by employing the ``extend matrix'', i.e.,
	\begin{equation}
	X:=\begin{pmatrix}0&\tilde{X}\\\tilde{X}^\dagger&0\end{pmatrix}\in\C^{(n+m)\times (n+m)},
	\end{equation} 
	which is square and hermitian. Naturally, if we can locate  the anchors of $X$, we can then easily obtain  the anchors of $\tilde{X}$. 
	
	Given a low-rank hermitian matrix $X\in\C^{(n+m)\times (n+m)}$, there exists an oracle, which can efficiently access the elements of $X$  by either performing an efficient computation of the matrix elements or authorizing access to a storage medium for the elements such as quantum RAM~\cite{giovannetti2008quantum}. With the oracle access to $X$, we could prepare a quantum circuit which simulates the unitary time evolution $e^{iXt}$ for a simulation time $t$, following the method introduced in~\cite{berry2007efficient,harrow2009quantum}.
	
	Note that, for an efficient quantum simulation, the input matrix is  required to be sparse. Thus, to simulate a dense matrix $X$ that may be encountered in SNMF, we  refer to the method  introduced in \cite{rebentrost2016quantum}. Specifically, $X$ can be efficiently embedded  into a large sparse matrix $S_{X}\in\mathbb{C}^{(n+m)^2\times (n+m)^2}$. Since $S_{X}$ is sparse, we can use $S_X$ to conduct an efficient quantum simulation. This  ``modified swap matrix"  $S_X$ is defined as
	\begin{equation}\label{eqn:S_X}
	S_{X}:=\sum_{i,j=1}^{n+m}X_{i,j}\ket{i}\bra{j}\otimes\ket{j}\bra{i}\in\C^{(n+m)^2\times (n+m)^2},
	\end{equation}  
	where $X_{i,j}$ is the $(i,j)$-th entry of $X$. It is easy to see that $S_{X}$ is one-sparse, at which the number of non-zero elements in each column and in each row  is one. 
	
	We show that, applying the simulated unitary time evolution $e^{iXt/(n+m)}$ to the target state $\sigma$, is equivalent to applying $e^{iS_{{X}} t}$ on $\rho\otimes \sigma$, where the ancillary state $\rho=\ket{\vec{1}}\bra{\vec{1}}$  and $\ket{\vec{1}}=\frac{1}{\sqrt{n+m}}\sum_i\ket{i}$. 
	Since $1/(n+m)$ is only a scale factor that does not influence the final result, $e^{iXt}$ can therefore be  efficiently simulated in quantum computer.
	We can trace out the  ancillary state $\rho$ from the quantum simulation system $e^{-iS_X\Delta t}\rho\otimes \sigma$ by
	\begin{eqnarray} \label{eqn:S_x=X}
	&&tr_{\rho}e^{iS_{{X}}\Delta t}\rho\otimes\sigma e^{-iS_{{X}}\Delta t} \nonumber\\
	&&=  e^{i\frac{X}{(n+m)}\Delta t}\sigma e^{-i\frac{X}{(n+m)}\Delta t} + O(\Delta t^2)~.
	\end{eqnarray}
	This indicates that for small time $\Delta t$, evolving with the modified swap matrix $S_X$ on the large system $\rho\otimes\sigma$ is equivalent to evolving with $X/(n+m)$ on the $\sigma$ system with a negligible error. Then generalizing to any simulation $t$, we can slice $t$ into multiple $\Delta t$. And in each $\Delta t$, a copy of $\rho$ is required. With an efficient oracle access to the matrix elements, we can simulate sparse matrix $S_X$ with a constant number of oracle calls and a negligible error~\cite{berry2007efficient,harrow2009quantum}.
	
	Random vectors used in DCA will be prepared as quantum states according to ~\cite{grover2002creating,harrow2009quantum}. Specifically, for a normalized vector $\beta=[\alpha_0,\cdots,\alpha_{(n+m)-1}]^\top\in\C^{(n+m)}$, if all of the entries as well as $\sum_{i=i_1}^{i_2}|\alpha_i|^2$ are efficiently computable, where  $i_2>i_1$ are any numbers in $\{1,\cdots,n+m\}$,  we can prepare the state $\ket{\beta}=\sum_{i=1}^{n+m}\alpha_i\ket{i}$ efficiently.

	\underline{Concluding remark 1.} In the state preparation step, given an oracle access to the elements of a low-rank and normalized rows hermitian matrix $X\in\C^{(n+m)\times (n+m)}$, there exists a quantum algorithm \cite{berry2007efficient,harrow2009quantum} which simulates the unitary time evolution $e^{-i{X}t/{(n+m)}}$ in runtime $O({\rm poly~\log}~(n+m))$. Likewise, given an oracle access to the elements of classical random and normalized vectors $\{\beta_i\}_{i=1}^s$, we can efficiently prepare corresponding quantum states $\{\ket{\beta_i}\}_{i=1}^s$ under a logarithmic runtime. 
	

	
	\subsection{Quantum Algorithm for Linear Operations}\label{sec:QNMF}
	
	
	After classical data are read into quantum forms, we shall utilize quantum principal component analysis scheme (QPCA) \cite{lloyd2014quantum} and its subsequent phase estimation algorithm~\cite{shor1999polynomial} to obtain a quantum state $\ket{\psi_i}$ which is proportional to random projections $X\beta_i$, for $i=\{1,\dots,s\}$. Let the eigen-decomposition of $X$ be $\sum_{j}\lambda_j\proj{u_j}$, where $\lambda_j$ and $\ket{u_j}$ stand for eigenvalues and their corresponding eigenvectors.
	Specifically, for the dense matrix case, with the oracle $\Lambda_{q}(\cdot)$, the exponential matrix $e^{iXt_0/(n+m)}$ with simulation time $t_0$ is applied to $\ket{\beta_i}$, resulting in
	\begin{equation}{\label{eqn:pea}}
	\begin{aligned}
	&\Lambda_{q}(e^{{iXt_0}/(n+m)})\ket{k}\ket{\beta_i}= \frac{1}{\sqrt{2^q}}\sum_{k}\ket{k}e^{ikXt_0/(n+m)}\ket{\beta_i}~, \nonumber
	\end{aligned}
	\end{equation}   
	where $q$ is a positive integer, $\ket{k}$ is composed with $q$ qubits (i.e., $\ket{k}=\ket{0}^{\otimes q}$) and the oracle $\Lambda_{q}(\cdot)$ applies $k$ times of  $e^{iXt_0}$ onto $\ket{\beta_i}$. 
	
	Next, taking $\ket{k}$ as the eigenvalue register with quantum operations,
	we can obtain the quantum state
	\begin{equation}
	\frac{1}{\sqrt{\sum_j|\beta_j|^2}}\sum_{\frac{|\lambda_j|}{(n+m)}\geq \epsilon}\beta_j\ket{\frac{\lambda_j}{(n+m)}}\ket{u_j}~,\nonumber
	\end{equation}
	where $\beta_j=\braket{u_j|\beta_i}$ in time $O(1/\epsilon)$.

	To obtain the analogous quantum form $X\ket{\beta_i}=\sum_{j}\lambda_j\braket{u_j|\beta_i}\ket{u_j}~,$
	that corresponds to the result of the random projection $X\beta_i$, the eigenvalues will be extracted into probability amplitudes of the resulting quantum state.
	We then follow the procedure in~\cite{harrow2009quantum}. Specifically, through introducing an ancilla qubit, applying rotating condition on $\ket{\lambda_j}$, and uncomputing the eigenvalue register, the resulting quantum state is proportional to
	\begin{eqnarray}
	&\sum_{\frac{|\lambda_j|}{(n+m)}\geq \epsilon}\beta_j\ket{u_j}&\left(\sqrt{1-\frac{\lambda_{j}^2}{C^2(n+m)^2}}\ket{0}\right.\nonumber\\
	&&\left.+\frac{\lambda_{j}}{C(n+m)}\ket{1}\right)~, \nonumber
	\end{eqnarray}
	where $C= O(1/\lambda_{max})$ and $\lambda_{max}$ is the largest eigenvalue of $X$.
	
	Measuring the last qubit, conditioned on seeing $1$ \cite{harrow2009quantum}, the final output state is proportional to $X\beta_i$, i.e.,
	\begin{equation}\label{eq: psi}
	\ket{\psi_i}=\frac{1}{\sqrt{\sum_j\frac{|\beta_j\lambda_j|^2}{C^2(n+m)^2}}}\sum_{j}\beta_j\frac{\lambda_{j}}{C(n+m)}\ket{u_j}.
	\end{equation}
	
	\underline{Concluding remark 2.} Given the quantum circuit that simulates $e^{iS_Xt}$ and a quantum state $\ket{\beta_i}$ encoding the vector $\beta_i$, there exists a quantum algorithm for linear operations that outputs a quantum state $\ket{\psi_i}$, c.f., equation~(\ref{eq: psi}), which is proportional to the vector $X\beta_i$, in time $\tilde{O}(1/\epsilon)$. Combined with the quantum circuit and the state preparation step in Subsection~\ref{sec:read_X}, the total runtime complexity of computing $X\beta_i$ for $i=\{1,\dots, s\}$ is ${O}(poly~\log(n+m)/\epsilon)$. Let the desired error be $1/\epsilon=O(poly~\log(n+m))$, the runtime complexity is ${O}(poly~\log(n+m))$.

	\subsection{ Heuristic Post-Selection }\label{sec:measurement}
	The resulting quantum state $\ket{\psi_i}$, which is proportional to the vector $X\beta_i$, is used to determine $\bar{A}^i$. The method to  extract the expected index is non-trivial. 
	In DCA, as the resulting vector is given explicitly after a random projection, we can pick up the indexes with the largest and the smallest entries in time $O(n+m)$. In the quantum setting, the entries of $X\beta_i$ are encoded into the probability amplitudes of $\ket{\psi_i}$. Reading out all probability amplitudes is exponential expensive. Even employing efficient tomography methods, such as compressed sensing~\cite{gross2010quantum} or sample optimal tomography \cite{haah2017sample},
	the  runtime complexity is $\tilde{O}(n+m)$. 
	Such a large cost  breaks the exponential speedup achieved in the random projection step.
	
	For the purpose of preserving the quantum advantages,  we devise an alternative heuristic method to obtain $\bar{A}^i$. The \textit{heuristic post-selection} is to find the index of an  anchor in the projected space (corresponding to the maximum probability amplitude of $\ket{\psi_i}$) under a logarithmic runtime by consuming
	$N$ copies of $\ket{\psi_i}$. Given $N$ copies of $\ket{\psi_i}$, the procedure to perform  the \textit{heuristic post-selection} is 
	\begin{enumerate}
		\item  measuring each copy $\ket{\psi_i}$ by the computational basis ;
		\item recording the most appearing index among the $N$ outputs as the index of an anchor in the projected space.
	\end{enumerate}
	
	The quantum state $\ket{\psi_i}$ contains $(n+m)$ superposition states which correspond to $(n+m)$ indexes, i.e.,~the possible measurement outcomes are $\{1,\dots,n+m\}$ and the probability of obtaining the index $k$ ($k\in\{1,\dots,n+m\}$) is given by $p_k=|\braket{k|\psi_i}|^2$. Among the $N$ outputs, the  most frequently appearing index corresponds to the index with the largest absolute amplitude with a high probability, which is also the index of an anchor in the projected space. 
	
	
	The probability of finding the index of an anchor is proportional to the number of quantum state copies $\ket{\psi_i}$, where the index corresponds to the maximum absolute amplitude of the quantum state $\ket{\psi_i}$. Then, a natural question is  how many copies are sufficient to determine the index of an anchor in the projected space with a high probability. The number of quantum state copies influences the runtime complexity of a quantum algorithm. Namely, using the computational basis to measure one copy of quantum state requires runtime complexity $O(1)$, and the measurement runtime by the computational basis is proportional to the number of copies. In the following, we show that, under a plausible assumption, only $N=O(poly~\log (n+m))$ copies of $\ket{\psi_i}$ are sufficient to determine the index of an anchor in the projected space with a high probability.

	\begin{thm}\label{thm}
		Let $D$ be a multinomial distribution. If ${\bf x}\sim D$, we assume $P({\bf x}=i)=p_i, i\in\{1,\cdots, N\}$, and $\sum_{i=1}^{N}p_i=1$. Let $x_1,\cdots,x_{N}$ be examples independently sampled from $D$ and $N_i$ be the number of examples taking value of $i$. Let $p_{max}=\max\{p_1,\cdots,p_{N}\}$ and $p_{secmax}=\max\{p_1,\cdots,p_{N}\}\setminus p_{max}$. If $p_{max}-p_{secmax}>2\sqrt{{2\log{(4N/\delta)}}/{N}}$, then, for any $\delta>0$, with a probability at least $1-\delta$, we have
		\begin{eqnarray}
		\arg\max_i\{N_i|1\leq i \leq N\} 
		=\arg\max_i\{p_i|1\leq i \leq N\}~.
		\end{eqnarray}
	\end{thm}
	In Theorem \ref{thm}, we have a plausible assumption $p_{max}-p_{secmax}>2\sqrt{{2\log{(4N/\delta)}}/{N}}$, which is easy to satisfy in practice. To achieve the exponential speedup, we could set $N=\log^2{(n+m)}$ and then we have
	\begin{equation} p_{max}-p_{secmax}>2\sqrt{{2\log{(4\log^2{(n+m)}/\delta)}}/{\log^2{(n+m)}}}~,
	\end{equation}
	which will converge to zero as $N$ goes to infinity. 
	This implies that given the above plausible assumption, by using the proposed \emph{heuristic post-selection} method, we could find the measured index corresponding to the maximum absolute amplitude of the resulting quantum state with a high probability. Here, the measured index also corresponds to the index of an anchor in the projected space. Recall that measuring $O(poly\log(n+m))$ quantum state copies by the computational basis implements \emph{heuristic post-selection} method.
	
	
	The proof of Theorem \ref{thm} is based upon the following Breteganolle-Huber-Carol inequality \cite{van1996weak}:
	\begin{thm} [Breteganolle-Huber-Carol inequality]
		Let $D$ be a multinomial distribution with $l$ events probabilities $p_1,\cdots,p_{l}$. Let $N_i$ be the number of event $i$ sampled from randomly sampled $N$ events. Then, for any $\delta>0$, the following inequality holds
		\begin{equation}
		P\left(\sum_{i=1}^{l}\left|\frac{N_i}{N}-p_i\right|\geq\lambda\right) \leq 2^{l}\exp{\left(\frac{-N\lambda^2}{2}\right)}~.
		\end{equation}	
		
		{\bf Proof of Theorem 1.} By utilizing the Breteganolle-Huber-Carol inequality, for any $j\in\{1,\cdots, N\}$, and $\delta>0$, we have
		\begin{eqnarray}
		&&P\left(\left|\frac{N_j}{N}-p_j\right|+\left|\sum_{i\neq j}\left(\frac{N_i}{N}-p_i\right)\right|\geq\lambda\right) \nonumber\\
		&&\leq 4\exp{\left(\frac{-N\lambda^2}{2}\right)}~.
		\end{eqnarray}
	\end{thm}
	Let $\delta=4\exp{\left(\frac{-N\lambda^2}{2}\right)}$. The above inequality implies that for any given $j\in\{1,\cdots, N\}$, with probability at least $1-\delta$, we have
	\begin{eqnarray}
	&\left|\frac{N_j}{N}-p_j\right|+\left|\sum_{i\neq j}\left(\frac{N_i}{N}-p_i\right)\right|\leq \sqrt{\frac{2\log{(4/\delta)}}{N}}~.
	\end{eqnarray}
	By using the union bound of probability, we have that for any $\delta>0$ and any $j\in\{1,\cdots, N\}$, with probability at least $1-\delta$, for the following inequality holds
	\begin{eqnarray}
	\left|\frac{N_j}{N}-p_j\right|\leq \sqrt{\frac{2\log{(4N/\delta)}}{N}}~.
	\end{eqnarray}
	Since   $p_{max}-p_{secmax}>2\sqrt{{2\log{(4N/\delta)}}/{N}}$, it can be easily verified that, with a probability at least $1-\delta$, there is only one value $N_j/N$ that is in the $\sqrt{{2\log{(4N/\delta)}}/{N}}$-neighborhood of $p_j$.
	We therefore conclude that  $\arg\max_i\{N_i|1\leq i\leq N\}=\arg\max_i\{p_i|1\leq i\leq N\}$. \hfill$\blacksquare$

	We also conduct  experiments to test the case without the plausible assumption. We empirically find that, the $O(poly~\log{(n+m)})$ measurements are sufficient to locate the index with the largest absolute amplitude at a very high probability. Given limited page length, we do not detail the procedure of the experiment. In a nutshell, we first generate the synthetic data in accordance with \cite{zhou2013divide}. And then, we  convert the result of each random projection into a probability distribution. Afterwards, the Monte Carlo simulation is introduced to sample examples from the distribution \cite{binder1993monte}. Finally, the statistical results indicate that, with the  sample size $O(poly~\log(n+m))$, the index with the largest entry  can be located with a high probability.

	After measuring polynomial logarithmic $N$ copies by the computational basis with runtime $O(poly\log(n+m))$, the most appearing index should be recorded among $N$ outputs, which can be obtained by a classical searching algorithm. The $\bar{A}^i$ then be determined with the runtime $O(poly\log(n+m))$.
	After applying the \textit{heuristic post-selection} onto $s$ sub-problems, the  $\{\bar{A}^i\}_{i=1}^s$ will be obtained. This achieves the divide step of QDCA. It is worth  noting that, different to the classical DCA,  QDCA only obtains the index with the largest absolute entry value. Therefore, the number of random vectors $\beta_i$ should be doubled. Since $s\ll \min(n, m)$, we have $2s\ll n+m$ and $s$ is still  $O(r\log r)$.
	
	\underline{Concluding remark 3.} {Supported by Theorem \ref{thm}, when the data size is $(n+m)$, $O(poly \log(n+m))$ random samples measured by computational basis are sufficient to locate the index of an anchor in a projected space with a high probability. Adding the runtime to search the most frequently appeared index, the runtime of this step is  $O(poly \log(n+m))$.}

	\subsection{The Classical Conquer Step}\label{sec:class}
	Via the \emph{heuristic post-selection} method, $\{\bar{A}^i\}_{i=1}^s$ are collected in the classical form. As analysis in Theorem \ref{thm}, non-anchors may be collected with probability at most $\delta$. Therefore, equation (\ref{def:conquer}) is applied to determine the indexes of anchors. This part is completed by a classical sorting algorithm. Through employing the sorting algorithm \cite{knuth1998section},  $s$ indexes of $\{\bar{A}^i\}_{i=1}^s$ can be sorted in  time $O(s\log{s})$.  Since $s=O(r\log{r})$, the runtime complexity is $O(poly\log(n+m))$. After sorting,  top $r$ indexes are selected as $R$, {which are most frequently appeared indexes in all $s$ sub-problems.} With the selected $R$, the decomposed matrix $X(R,:)$ is obtained. 
	
	\underline{Concluding remark 4.} {Through employing the sorting algorithm on $\{\bar{A}^i\}_{i=1}^s$, the indexes of anchors are obtained with runtime complexity $O(poly\log(n+m))$.}	
	
	\section{Conclusion}\label{sec:conclusion}
	This paper presents QDCA to dramatically reduce the runtime to achieve the exponential speedup for extracting part-based representations from a large-scale matrix. Analogous to DCA, QDCA can also solve near-separable non-negative matrix factorization problem with the exponential speedup. 
	Moreover, the strategy of combining quantum and classical computations paves a new way to develop quantum machine learning algorithms. Through employing \emph{heuristic post-selection} method, the quantum advantage is achieved after reading out quantum data into the classical form. In the future, we plan to apply this QDCA scheme to various machine learning algorithms to achieve the quantum speedup.

\bibliographystyle{named}
\bibliography{1}
\end{document}